\documentclass[PRD,reprint,twocolumn,showpacs,
amssymb, amsmath, aps, showpacs, nofootinbib, superscriptaddress]{revtex4}

\usepackage{multirow}
\usepackage{hyperref}
\usepackage{graphicx}
\usepackage{dcolumn}
\usepackage{bm}
\usepackage{threeparttable}
\usepackage{subfig}
\usepackage{color,txfonts}
\usepackage{ulem}

\def\dif{\rm d}
\def\beq{\begin{equation}}
\def\eeq{\end{equation}}
\def\bey{\begin{eqnarray}}
\def\eey{\end{eqnarray}}

\def\lsim{\mathrel{\raise.3ex\hbox{$<$\kern-.75em\lower1ex\hbox{$\sim$}}}}
\def\gsim{\mathrel{\raise.3ex\hbox{$>$\kern-.75em\lower1ex\hbox{$\sim$}}}}

\begin{document}

\title{Hearing the birth of the lightest intermediate mass black holes with the second generation gravitational wave detectors}
\affiliation{Key Laboratory of Dark Matter and Space Astronomy, Purple Mountain Observatory, Chinese Academy of Sciences, Nanjing 210008, China}
\affiliation{University of Chinese Academy of Sciences, Yuquan Road 19, Beijing 100049, China}
\affiliation{TianQin Research Center for Gravitational Physics, Sun Yat-sen University, Zhuhai, 519082, China}
\affiliation{School of Astronomy and Space Science, University of Science and Technology of China, Hefei, Anhui 230026, China}
\author{Yun-Feng Liang}
\affiliation{Key Laboratory of Dark Matter and Space Astronomy, Purple Mountain Observatory, Chinese Academy of Sciences, Nanjing 210008, China}
\affiliation{University of Chinese Academy of Sciences, Yuquan Road 19, Beijing 100049, China}
\author{Yuan-Zhu Wang}
\affiliation{Key Laboratory of Dark Matter and Space Astronomy, Purple Mountain Observatory, Chinese Academy of Sciences, Nanjing 210008, China}
\affiliation{University of Chinese Academy of Sciences, Yuquan Road 19, Beijing 100049, China}
\author{Hao Wang}
\affiliation{Key Laboratory of Dark Matter and Space Astronomy, Purple Mountain Observatory, Chinese Academy of Sciences, Nanjing 210008, China}
\affiliation{University of Chinese Academy of Sciences, Yuquan Road 19, Beijing 100049, China}
\author{Xiang Li}
\email{xiangli@pmo.ac.cn}%
\affiliation{Key Laboratory of Dark Matter and Space Astronomy, Purple Mountain Observatory, Chinese Academy of Sciences, Nanjing 210008, China}
\author{Yi-Ming Hu}
\email{huyiming@mail.sysu.edu.cn}%
\affiliation{TianQin Research Center for Gravitational Physics, Sun Yat-sen University, Zhuhai, 519082, China}
\email{yiming.hu@aei.mpg.de (YMH)}%
  \author{Zhi-Ping Jin }
\affiliation{Key Laboratory of Dark Matter and Space Astronomy, Purple Mountain Observatory, Chinese Academy of Sciences, Nanjing 210008, China}
\affiliation{School of Astronomy and Space Science, University of Science and Technology of China, Hefei, Anhui 230026, China}
\author{Yi-Zhong Fan}
\email{yzfan@pmo.ac.cn}%
\affiliation{Key Laboratory of Dark Matter and Space Astronomy, Purple Mountain Observatory, Chinese Academy of Sciences, Nanjing 210008, China}
\affiliation{School of Astronomy and Space Science, University of Science and Technology of China, Hefei, Anhui 230026, China}
\author{En-Wei Liang}
\affiliation{Guangxi Key Laboratory for the Relativistic Astrophysics, Department of Physics,
Guangxi University, Nanning 530004, China}
\author{Da-Ming Wei }
\affiliation{Key Laboratory of Dark Matter and Space Astronomy, Purple Mountain Observatory, Chinese Academy of Sciences, Nanjing 210008, China}
\affiliation{School of Astronomy and Space Science, University of Science and Technology of China, Hefei, Anhui 230026, China}


\begin{abstract}
  Intermediate mass black holes (IMBHs) with a mass between $10^{2}$ and $10^{5}$ times that of the sun, which bridges the {mass gap between the} stellar-mass black holes and the supermassive black holes, are crucial in understanding the evolution of the black holes. Although they are widely believed to exist, decisive evidence has long been absent. Motivated by the successful detection of massive stellar-mass black holes by advanced LIGO, through the gravitational wave radiation during the binary merger, in this work we investigate the prospect of detecting/identifying the lightest IMBHs (LIMBHs; the black holes $\gtrsim 100M_\odot$) with the second generation gravitational wave detectors. In general, the chance of hearing the birth of the LIMBHs is significantly higher than that to identify pre-merger IMBHs. The other formation channel of LIMBHs, where stars with huge mass/low-metallicity directly collapse, is likely ``silent", so the merger-driven birth of the LIMBHs may be the only ``observable" scenario in the near future.  Moreover, the prospect of establishing the presence of (lightest) intermediate mass black holes in the O3 run and beyond of advanced LIGO is found quite promising, implying that such an instrument could make another breakthrough on astronomy in the near future. The joining of other detectors like advanced Virgo would only increase the detection rate.
\end{abstract}


\maketitle
\section {Introduction}

In the Universe a large amount of black holes (BHs) have been identified in astronomical observations \cite{Fender2012,Volonteri2012}. So far about two dozens of BHs of stellar mass (i.e., the masses are below $\sim 100M_\odot$) have been detected in X-ray binaries within the Milky Way and a few nearby galaxies \cite{Farr2011} and the heaviest has a mass of $\sim 20~M_\odot$. The stellar-mass BHs are expected to form when very massive stars collapse at the end of their life cycle \cite{Woosley2002} or some neutron star binaries merge. After its formation, a BH can continue to grow by the mass accretion from its surroundings. Another major type of the observed BHs is the so-called supermassive black hole (SMBH), which appears in the center of, for example, the Milky Way and almost all the massive galaxies. The masses of SMBHs are on the order of $10^5-10^9~M_\odot$ \cite{Antonucci1993}. The existence of intermediate-mass black holes (IMBH), with masses of $10^2-10^5~M_\odot$ that bridges stellar-mass and super-massive
BHs thus is very helpful in understanding the evolution of the black holes, has not been firmly established though there are some intriguing candidates \cite{Volonteri2012,Kiziltan2017,Pasham2014}.

In this work we focus on the birth of the lightest IMBHs that in principle could be formed either by the direct collapse of the most massive stars with extremely-low metallicity (also known as the first stars) or the merger of two massive stellar-mass BHs. The direct collapse of the first stars, however, are widely expected to be silent/distant and hard to monitor \cite{Woosley2002}. Hence in the foreseeable future, the detectable channel for the birth of the LIMBHs is likely to be the merger of two massive stellar-mass black holes.
The merger rate of binary black holes (BBH) was expected to be high in some models \cite{Tutukov1993}.
The first gravitational wave (GW) event was from the merger of a pair of black holes with the masses of 25 $M_\odot$ and 36 $M_\odot$, respectively \cite{2016PhRvL.116f1102A}.
The newly formed black hole has a mass of $\sim 62~M_\odot$, which is the heaviest stellar-mass black hole recorded by far.
The six pre-merger BHs identified in current coalescence events/candidates can be simply fitted by a single power-law distribution and yields a mass function of ${\rm d}N/{\rm d}m_{\rm BH} \propto m_{\rm BH}^{-2.5^{+1.5}_{-1.6}}$ up to $m_{\rm BH}\geq 36M_\odot$ \cite{Abbott2016c,Abbott2016d}. With a high rate density $\sim 10^{2}~{\rm Gpc^{-3}~yr^{-1}}$ of BH binary mergers, as inferred from the O1 run data and implied by the ongoing O2 run (http://www.ligo.org/news/) of advanced LIGO (aLIGO), the second generation gravitational wave detectors are expected to detect hundreds of such events per year in 2020s \citep{Abbott2016d,2016arXiv161101157K}.
The main purpose of this work is thus to investigate the prospect of ``hearing" the birth of the LIMBHs with the second-generation gravitational wave detectors in particular the aLIGO and advanced Virgo (AdV) \cite{Abbott2016LRR}.

\section {Method}
We concentrate on IMBHs in compact binary coalescence systems.
In this work we focus on two issues: one is the ratio between the detection rates of newly-formed LIMBHs ($R_{\rm birth}$) and the coalescence of compact binaries involving IMBHs ($R_{\rm pre-merg}$), defined as $r_{\rm imbh}\equiv R_{\rm birth}/R_{\rm pre-merg}$; the other is the prospect of hearing the birth of the LIMBHs with the advanced LIGO.

In \cite{Abbott2016d}, two ``envelope" distributions for BBH
masses were assumed, including (i) ``a distribution uniform (flat) over the logarithm of
component masses (i.e., $p(m_{1},m_{2})\propto m_{1}^{-1}m_{2}^{-1}$)"
and (ii) ``a power-law distribution in the primary
mass, $p(m_1)\propto m_1^{-2.35}$, with a uniform distribution
on the second mass". The additional constraints are $m_1\geq m_2\geq 5~M_{\odot}$, i.e., a mass gap is assumed between the maximum gravitational mass of neutron stars ($\sim 2M_\odot$)
and the lightest BHs ($m_{\rm gap}=5M_\odot$), as motivated by the X-ray binary observations/modeling \cite{Farr2011}. In reality, the
first distribution likely overestimates the fraction of
high-mass black holes while the second, resembling the initial
mass function for stars \cite{Salpeter1955}, plausibly overestimates the fraction of low-mass
black holes \cite{Abbott2016d}. Since this work concentrates on the mergers of massive stellar-mass black holes and aims to get the conservative estimate on the formation of LIMBHs,
in the following approach we take the second mass distribution model.
For the same concern, an additional exponential cutoff at the high end of the BH mass function is assumed. Therefore, following \cite{2016arXiv161101157K,WangH2017}, the BH mass functions (i.e., $P_1({m_1})$ and $P_2({m_2})$) are taken as
\begin{equation}\label{M1MF}
 P_1\left( {{m_{\rm{1}}}} \right) \propto m_{\rm{1}}^{ - \alpha }{e^{ - {m_{\rm{1}}}/{m_{{\rm{cap}}}}}},\ {{\rm for}\ m_1>m_{\rm gap}},
\end{equation}
\begin{equation}\label{M2MF}
 P_2({m_{\rm{2}}}) \propto {({m_{\rm{2}}}/{m_{\rm{1}}})^\beta },\ {{\rm for}\ m_{\rm gap}<m_2<m_1},
\end{equation}
where $m_{\rm{cap}}$ is the possible high-mass cutoff of the mass function, as mentioned above, which is however not well constrained yet. Nevertheless, motivated by the detection of
a pre-merger BH as massive as $\sim 36M_\odot$ in GW150914, it is reasonable to assume that $m_{\rm cap}\geq 40M_\odot$.
In the following analysis, we simply assume $40M_\odot \leq m_{\rm cap} \leq 100M_\odot$. In some literature it is assumed that the BH mass function resembles that of the massive stars, i.e., $\alpha=2.35$ \cite{Abbott2016d,2016arXiv161101157K}.
We take it as the fiducial value while other values of $\alpha$ between 2 and 3 are also investigated in estimating $r_{\rm imbh}$.
Following \cite{Abbott2016d,2016arXiv161101157K} we simply take $\beta=0$ throughout this work.

The expected distribution that describes the masses of the binary members in the observed GW events is \cite{2016arXiv161101157K,WangH2017}
\begin{eqnarray}
\frac{{\dif}N}{{\dif}m_1{\dif}m_2} &=& P_1({m_1})P_2({m_2}){\int_0^{z_{\rm max}}{\frac{{\cal R}_{\rm BBH}{\cdot}}{(1+z)}}\frac{dV_z}{dz}dz},
\label{eq:dN}
\end{eqnarray}
where ${\cal R}_{\rm BBH}$ is the merger rate of BBHs,
$(1+z)$ accounts for the difference in clock rates at the merger and at the detector and $V_z$ denotes the comoving volume of the space within which the GW events can be detected by LIGO, i.e.,
${dV_z}/{dz}={4\pi{c}{\cdot}d_c^2(z)}/{H(z)}$,
where $d_c$ is the comoving distance of the GW event, $c$ is the speed of light and $H(z)$ is the hubble parameter.
The maximum redshift $z_{\rm max}$ that LIGO can probe depends on both $m_1$ and $m_2$, which is described in more details as below.

The optimized oriented signal-to-noise ratio of a GW event with given pre-merger BHs masses can be calculated as \cite{1994PhRvD..49.2658C}
\begin{equation}
\rho_{\rm opt}^2=4\int_{0}^{\infty}{\frac{|\tilde{h}(f)|^2}{S_n(f)}df},
\end{equation}
where $\tilde{h}(f)$ refers to the Fourier Transformation of a GW strain at the detector. Here we use a general form of an angle-averaged $(S/N)=\langle\rho \rangle$ expressed as \cite{1994PhRvD..49.2658C}
\begin{equation}
({S/N})^{2} = \frac{4}{5} \int_{f_{\rm min}}^{f_{\rm max}}
df \frac{h_{c}^{2}(f)}{S_{n}(f) (2 f)^{2}},
\label{eq:SNR}
\end{equation}
where $h_c(f)$ is the observed strain amplitude.
Generally, we assume a GW event is ``detected" if its $S/N$ is greater than 8.
For the strain noise amplitude of aLIGO in full design sensitivity, $S_n(f)=h^2_n(f)$, we adopt the analytical form in \cite{Ajith2011}. For O2 and O3, we use the upper boundaries of the sensitivity curves in \cite{Abbott2016LRR} (their Figure 1) to get a conservative estimation on the event rates.
For the observed strain amplitude, it can be expressed as \cite{Cholis2016}
\begin{equation}
h_c(f_{\rm obs})=\sqrt{2}\frac{1+z}{\pi{d_L(z)}}\sqrt{\frac{{\dif}E}{{\dif}f_{\rm s}}},
\label{eq:obs_amp}
\end{equation}
 with ${\dif}E/{{\dif}f_{\rm s}}$ denoting the spectral energy density. Both inspiral phase and merger phase of the coalescence are considered. We follow the spectral energy densities given by \cite{Cholis2016} (their equation (26) and (30)) as well as their assumptions and parametrization.
With the above equations we can determine the $z_{\rm max}$ that LIGO can reach for a given $m_1$ and $m_2$.

The distribution of the total masses of the pre-merger BHs is governed by
\beq
\frac{{\dif}N}{{\dif}m_{\rm tot}} \propto \int_{m_{\rm gap}}^\frac{m_{\rm tot}}{2}\left[{P_1({m_{\rm tot}-m_2})P_2({m_2}){\int_0^{z_{\rm max}}{\frac{{\cal R}_{\rm BBH}{\cdot}}{(1+z)}}\frac{dV_z}{dz}dz}}\right]{\dif}m_2.
\label{eq:m3_distr}
\eeq
As a result of the gravitational wave radiation, the nascent BH's mass $m_3$ will be less than $m_{\rm tot}=m_1+m_2$. We use $\eta$ to denote the fraction of the mass radiated away by GW. In a specific merger event $\eta$ depends on the total spin of the system along the orbital angular momentum, the mass ratio and the asymmetric mass ratio between the two BHs etc \cite{Barausse2012,Husa2016}. These necessary information, unfortunately, are largely uncertain. Nevertheless, in the strongest gravitational wave radiation scenarios, the fraction of the energy carried away is $\eta \sim 10\%$ \cite{Barausse2012,Husa2016}. For GW150914, GW151226 and LVT151012, about $0.04-0.05$ fraction of total masses were radiated away as gravitational waves \cite{Abbott2016d}. Motivated by these facts, we simply assume $\eta=0.05$ and the mass of the nascent BH formed in the merger can be straightforwardly evaluated (i.e., $m_3=(1-\eta)m_{\rm tot}$).

In the model of a power-law distribution in the primary
mass, the inferred BBH coalescence rate with the O1 run data is $97_{-67}^{+135}~{\rm Gpc^{-3}yr^{-1}}$ (90\% confidence level) for $\alpha=2.35$ and $m_{\rm tot}\leq 100M_\odot$ \cite{Abbott2016d}. Since our current model is rather similar except the introduction of $m_{\rm cap}(\geq 40M_\odot)$, it is reasonable to assume that ${\cal R}_{\rm BBH}\sim 30-232~{\rm Gpc^{-3}yr^{-1}}$ (90\% confidence level). The results of the ongoing O2 run have not been published yet but several candidates/events have been reported in the LSC news (http://www.ligo.org/news/), implying that ${\cal R}_{\rm BBH}\sim 100~{\rm Gpc^{-3}yr^{-1}}$ is reasonable. Following \cite{Abbott2016d} in this work no dependence of the merger rate on redshift is assumed. If ${\cal R}_{\rm BBH}$ increases with redshift for $z\leq 1-2$, the detection prospect of LIMBHs estimated in this work is {\it conservative}.

\section{Result}
\label{result}
\begin{figure}[!tbp]
\centering
\includegraphics[width=0.6\textwidth]{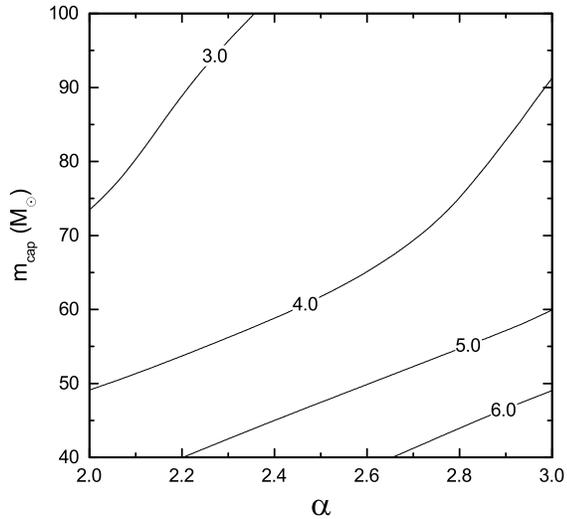}
\caption{The ratio between the detection rates of newly-formed LIMBHs and the coalescence of compact binaries involving IMBHs ($r_{\rm imbh}$) as a function of $\alpha$ and $m_{\rm cap}$ for $\eta=0.05$ (i.e., the black sold line). The value of each $r_{\rm imbh}$ line is marked in the plot.}
\label{fig:ratio}
\end{figure}

In Fig.\ref{fig:ratio} we present the ratio between the detection rates of newly-formed LIMBHs and the coalescence of compact binaries involving IMBHs $r_{\rm imbh}$ for $\eta=0.05$. The general trend is that the harder the mass function (i.e., with smaller $\alpha$ and higher $m_{\rm cap}$), the smaller the $r_{\rm imbh}$. This is anticipated since in such cases there is a higher possibility to get the pre-merger IMBHs. Nevertheless, in general, we have $r_{\rm imbh}\geq 3$ (for the soft BH mass function, $r_{\rm imbh}\sim 10$ is possible), implying that the first LIMBH identified in the gravitational wave data is very likely the nascent one formed in the mergers (i.e., $m_3\geq 100M_\odot$) rather than the massive compact object involved in the merger (i.e., $m_1\geq 100M_\odot$). But since $r_{\rm imbh}$ is just modestly high, it is still possible that in the first gravitational wave event ``involving" IMBH we have both $m_1\geq 100M_\odot$ and $m_3\geq 100M_\odot$.

\begin{figure}[!tbp]
    \centering
\includegraphics[width=0.55\textwidth]{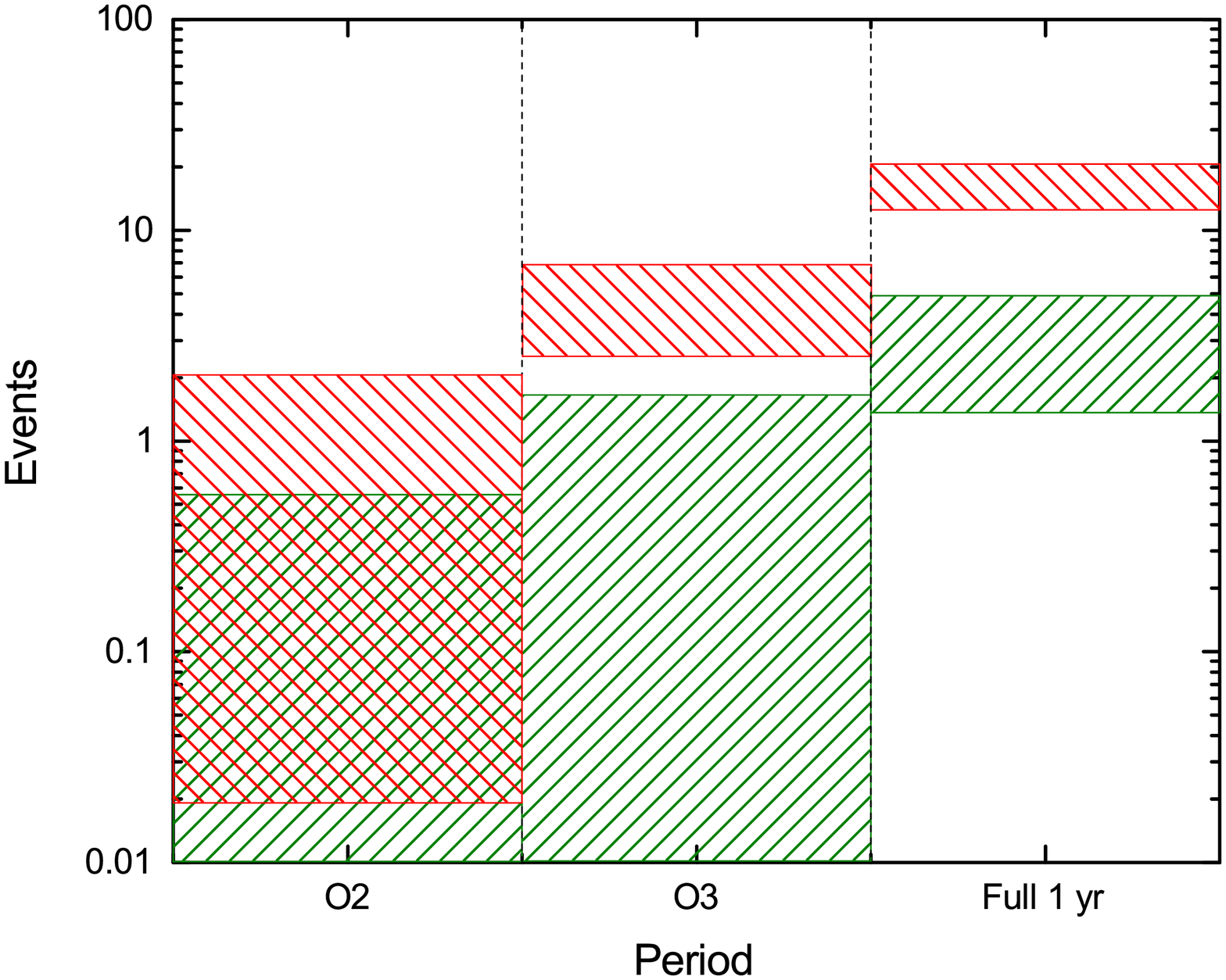}
  \caption{The prospects of hearing the birth of LIMBHs (shaded regions in red) or detecting the pre-merger LIMBHs (shaded regions in green) in the runs of O2, O3 and full design sensitivity (1 yr) of aLIGO detectors, respectively. The physical parameters are adopted as $\alpha=2.35$, $m_{\rm cap}=40M_\odot$, $\eta=0.05$, $f_{\rm dc}=0.8$ and ${\cal R}_{\rm BBH}=100~{\rm Gpc^{-3}yr^{-1}}$.  The Poisson noise is adopted in estimating the allowed event number regions (68\% confidence level).}\label{fig:detection}
\end{figure}

With the given initial mass functions of the pre-merger BHs, $\eta$ and $R_{\rm BBH}$, it is straightforward to estimate the fraction of mergers that can yield IMBHs (i.e., $f_{\rm imbh}$). For a given network, the time to see
its first event can be estimated as following. It is described by a Poisson process, with
the probability distribution of waiting a time $\tau$ before detecting the first
event given by $e^{-\tau f_{\rm imbh}R_{\rm BBH}V}$, for an event rate $R_{\rm BBH}$. We define $\tau_{\rm first,eff}=2.3/f_{\rm imbh}R_{\rm BBH}V$ as
the effective waiting time by which, in 90\% of cases, the first event
will have been observed. Right now the average reach of the aLIGO network for binary merger events has been around 700 Mpc for 30+30 Msun mergers. Hence we simply assume that $V\sim 1~{\rm Gpc^{3}}$. For $f_{\rm imbh} \sim 0.05-0.1$, we have $t_{\rm first}=\tau_{\rm first,eff}/f_{\rm dc}^{2}\approx 0.8~{\rm yr}~(f_{\rm imbh}/0.05)(R_{\rm BBH}/100{\rm Gpc^{-3}~{\rm yr}^{-1}})(V/1{\rm Gpc^{-3}})(f_{\rm dc}/0.8)^{-2}$, where $f_{\rm dc}\sim 0.8$ is the factor of the duty cycle for each detector of aLIGO. The O2 run of aLIGO will last about six months. Therefore the successful detection of LIMBH in O2 run of aLIGO is possible but not guaranteed. The O3 run of aLIGO is anticipated to last 9 months with the further-improved sensitivity, for which the chance of firmly detecting LIMBH is high. In Fig.\ref{fig:detection} we present the anticipated rate for $\alpha=2.35$ and $m_{\rm cap}=40M_\odot$ (a larger $m_{\rm cap}$ will further increase the detection rate of LIMBHs) and one can see that the detection prospect is quite promising in particular in O3 run or the first year full run of the aLIGO, which are reasonably in agreement with our analytical estimate.  {Note that even for a low ${\cal R}_{\rm BBH}\sim 30~{\rm Gpc^{-3}yr^{-1}}$, the birth of LIMBHs is expected to be directly heard during the first year full-run of aLIGO.} Where the target strain sensitivities for O2, O3 are taken from \cite{Abbott2016LRR} (see Fig.1 therein) and the full-sensitivity runs from \cite{Ajith2011} \cite{Abbott2016LRR}. In Fig.\ref{fig:detection} we have just estimated the events expected for the aLIGO detectors and the joining of other detectors like AdV, KAGRA and LIGO-India would only enhance the detection prospect.

\section {Summary}
In this work we have estimated the ratio between the detection rates of the formation of the IMBHs in the mergers and that of pre-merger IMBHs (note that these BHs are expected to be among the lightest IMBHs since their masses are expected just slightly above $100M_\odot$) and the prospect of hearing the birth of the lightest IMBHs with the second generation gravitational wave detectors in particular the aLIGO. The chance of ``hearing" the birth of a IMBH is found to be higher than identifying a pre-merger IMBH, as expected, though the possibilities are only different by a factor of $\sim 10$ or smaller (see Fig.\ref{fig:ratio}). Nevertheless, the prospect of successful detection of the lightest IMBHs is quite promising. In the optimistic case such objects will be detected in the O2 run of aLIGO (see Fig.\ref{fig:detection}). Quite a few IMBHs may be detectable in the upcoming O3 run of aLIGO detectors {(note that in Fig.\ref{fig:detection} we take ${\cal R}_{\rm BBH}\sim 100~{\rm Gpc^{-3}yr^{-1}}$. A more conservative assumption of  ${\cal R}_{\rm BBH}\sim 30~{\rm Gpc^{-3}yr^{-1}}$ will not change the conclusion qualitatively).}
The Cosmos will deliver more wonders through the channel of gravitational wave, and we suggest that in addition to the possible detection of mergers of NS-BH and/or NS-NS binaries \cite{Clark1977,LiX2017,Abbott2017}, in the dawn of GW astronomy era (i.e., the O2 and O3 runs), the lightest IMBHs will be plausibly detected. Hence the existence of (lightest) IMBHs will be firmly established though more massive ones are still to be robustly identified possibly with electromagnetic data until the performance of the third generation gravitational wave detectors. Though the direct collapse of very-massive stars with extremely-low metallicity may also produce the LIMBHs, such events are likely hard to detect. Therefore, the birth of the LIMBHs is expected to be only hearable/detectable in gravitational wave in the next decade. If instead there is still no any LIMBH recorded by aLIGO after its first year full-sensitivity run, the black hole mass function with a very sharp cutoff at $\sim 40M_\odot$ will be favored or alternatively the typical mass ratio $(q\equiv m_2/m_1)$ should be significantly smaller than 1.


\acknowledgments  This work was supported in part by 973 Programme of China (No. 2013CB837000 and No. 2014CB845800), by NSFC under grants 11525313 (the National Natural Fund for Distinguished Young Scholars), 11273063 and 11433009, by the Chinese Academy of Sciences via the External Cooperation Program of BIC (No. 114332KYSB20160007).



\begin{thebibliography}{}

\bibitem[Fender \& Belloni (2012)]{Fender2012} R. Fender, and T. Belloni, Science {\bf 337}, 540-544 (2012).

\bibitem[Volonteri (2012)]{Volonteri2012} M. Volonteri,
Science {\bf 337}, 544-547 (2012).

\bibitem[Farr et al. (2011)]{Farr2011} W. M. Farr, N. Sravan, A. Cantrell, et al. Astrophys. J., {\bf 741}, 103 (2011).

\bibitem[Woosley \& Heger (2002)]{Woosley2002} S. E. Woosley and A. Heger, Rev. Mod. Phys., {\bf 74}, 1015 (2002).

\bibitem[Antonucci (1993)]{Antonucci1993} R. Antonucci,
Ann. Rev. Astron. Astrophys. {\bf 31}, 473-521 (1993).

\bibitem[Kiziltan et al. (2017)]{Kiziltan2017} B. K$\iota$z$\iota$ltan, H. Baumgardt and A. Loeb, Nature, {\bf 542}, 203 (2017).

\bibitem[Pasham et al. (2014)]{Pasham2014} D. R. Pasham, T. E. Strohmayer and R. F. Mushotzky, Nature, {\bf 513}, 74 (2014).

\bibitem[Tutukov \& Yungelson (1993)]{Tutukov1993} A. V. Tutukov and L. R. Yungelson,
 Mon. Not. R. Astron. Soc. {\bf 260}, 675 (1993);
V. M. Lipunov, K. A. Postnov and M. E. Prokhorov,
Mon. Not. R. Astron. Soc. {\bf 288},
245 (1997);
R. Voss and T. M. Tauris,
Mon. Not. R. Astron. Soc. {\bf 342}, 1169 (2003);
M. Dominik, E. Berti, R. O¡¯Shaughnessy, I. Mandel, K.
Belczynski, C. Fryer, D. E. Holz, T. Bulik and F. Pannarale,
Astrophys. J. {\bf 806}, 263 (2015);
C. L. Rodriguez, M. Morscher, B. Pattabiraman, S. Chatterjee, C.-J. Haster and F. A. Rasio,
Phys. Rev. Lett. {\bf 115}, 051101 (2015).

\bibitem[Abbott et al.(2016a)]{2016PhRvL.116f1102A} B.~P. Abbott, R. Abbott, T.~D. Abbott, et al. (LIGO Scientific Collaboration and Virgo Collaboration) \ Phys. Rev. Lett., {\bf 116}, 061102 (2016).

\bibitem[Abbott et al.(2016b)]{Abbott2016c} B.~P. Abbott, R. Abbott, T.~D. Abbott, et al. (LIGO Scientific Collaboration and Virgo Collaboration) \ Phys. Rev. Lett., {\bf 116}, 241103 (2016).

\bibitem[Abbott et al.(2016c)]{Abbott2016d} B. P. Abbott, R. Abbott, T. D. Abbott, et al. (LIGO Scientific Collaboration and Virgo Collaboration) \ Phys. Rev. X, {\bf 6}, 041015 (2016).

\bibitem[Kovetz et al.(2016)]{2016arXiv161101157K} E.~D. Kovetz, I. Cholis, P.~C. Breysse \& M. Kamionkowski,\ 2016, arXiv:1611.01157


\bibitem[Abbott et al.(2016d)]{Abbott2016LRR} Abbott, B. P., Abbott, R., Abbott, T. D., et al. Liv. Rev. Relativ., {\bf 19}, 1 (2016)


\bibitem[Salpeter (1955)]{Salpeter1955} E. E. Salpeter, Astrophys. J. {\bf 121}, 161 (1955).

\bibitem[Wang et al.(2017)]{WangH2017} H. Wang, Y. Z. Wang, Y. F. Liang, et al.,\ 2017, ApJ submitted

\bibitem[Flanagan \& Hughes(1998)]{Flanagan1998} {\'E}.~{\'E}. Flanagan and S.~A. Hughes, \prd, {\bf 57}, 4535 (1998).

\bibitem[Cutler \& Flanagan(1998)]{1994PhRvD..49.2658C} C. Cutler \& E. Flanagan, \prd, {\bf 49}, 2658 (1994).

\bibitem{Cholis2016} I.~Cholis, E.~D.~Kovetz, Y.~Ali-Ha\"{i}moud, S.~Bird, M.~Kamionkowski, J.~B.~Mu\~{n}oz and A.~Raccanelli, \prd, {\bf 94}, 084013 (2016).

\bibitem[Ajith(2011)]{Ajith2011} P. Ajith, \prd, {\bf 84}, 084037 (2011).

\bibitem[Barausse (2012)]{Barausse2012} E. Barausse, V. Morozova \& L. Rezzolla, Astrophys. J. {\bf 758},  63 (2012).

\bibitem[Husa (2016)]{Husa2016} S. Husa et al., Phys. Rev. D, {\bf 93}, 044006 (2016).

\bibitem[Clark \& Eardley (1977)]{Clark1977} J. P. A. Clark and D. M. Eardley, Astrophys. J., {\bf 215}, 311 (1977).

\bibitem[Li et al.(2017)]{LiX2017} X. Li, Y. M. Hu, Z. P. Jin, Y. Z. Fan \& D. M. Wei, 2017, arXiv:1611.01760

\bibitem[Abbott et al.(2017)]{Abbott2017} B. P. Abbott, R. Abbott, T.~D. Abbott et al. (LIGO Scientific Collaboration and Virgo Collaboration). Astrophys. J. Lett. {\bf 832}, L21 (2017).



\end{thebibliography}
\end{document}